\documentclass[aip,graphicx,twocolumn, 10 pt]{revtex4-1}

\usepackage{graphicx}

\usepackage{xcolor}
\usepackage{amsmath}

\newcommand*{\citen}[1]{%
  \begingroup
    \romannumeral-`\x 
    \setcitestyle{numbers}%
    \cite{#1}%
  \endgroup   
}

\draft 

\begin{document}


\title{Determination of the dynamic Young's modulus of quantum materials in piezoactuator-driven uniaxial pressure cells using a low-frequency a.c. method} 



\author{Caitlin I. O’Neil$^{1,2}$}
\email[]{caitlin.oneil@cpfs.mpg.de}
\noaffiliation

\author{Zhenhai Hu$^{1,2}$}
\noaffiliation

\author{Naoki Kikugawa$^{3}$}
\noaffiliation

\author{Dmitry A. Sokolov$^{1}$}
\noaffiliation

\author{Andrew P. Mackenzie$^{1,2}$}
\noaffiliation

\author{Hilary M. L. Noad$^{1}$}
\noaffiliation

\author{Elena Gati$^{1}$}
\email[]{elena.gati@cpfs.mpg.de}
\noaffiliation

\affiliation{Max Planck Institute for Chemical Physics of Solids, Dresden, Germany}
\affiliation{Scottish Universities Physics Alliance, School of Physics and Astronomy, University of St Andrews, St Andrews, UK}
\affiliation{National Institute for Materials Science, Tsukuba, Japan}


\date{\today}

\begin{abstract}
We report on a new technique for measuring the dynamic Young's modulus, $E$, of quantum materials at low temperatures as a function of static tuning strain, $\epsilon$, in piezoactuator-driven pressure cells. In addition to a static tuning of stress and strain, we apply a small-amplitude, finite-frequency a.c. (1\,Hz$\,\lesssim\,\omega\,\lesssim\,$1000\,Hz) uniaxial stress, $\sigma_{ac}$, to the sample and measure the resulting a.c. strain, $\epsilon_{ac}$, using a capacitive sensor to obtain the associated modulus $E$. We demonstrate the performance of the new technique through proof-of-principle experiments on the unconventional superconductor Sr$_2$RuO$_4$, which is known for its rich temperature-strain phase diagram. In particular, we show that the magnitude of $E$, measured using this a.c. technique at low frequencies, exhibits a pronounced nonlinear elasticity, which is in very good agreement with previous Young's modulus measurements on Sr$_2$RuO$_4$ under [1\,0\,0] strain using a d.c. method (Noad \textit{et al.}, Science \textbf{382}, 447-450 (2023)). By combining the new a.c. Young's modulus measurements with a.c. elastocaloric measurements in a single measurement, we demonstrate that these a.c. techniques are powerful in detecting small anomalies in the elastic properties of quantum materials. Finally, using the case of Sr$_2$RuO$_4$ as an example, we demonstrate how the imaginary component of the modulus can provide additional information about the nature of ordered phases.
\end{abstract}


\maketitle 

\section{Introduction}
\label{sec:intro}

Recent years have witnessed a surge in the study of elastic properties of quantum materials, driven by the discovery of novel collective electronic phases that exhibit a strong coupling to the underlying crystal lattice. A prominent example is the observation of nematicity in a number of unconventional superconductors \cite{Fernandes14,Hinkov08}. Here, measurements of the elastic constants have revealed a huge lattice softening \cite{Boehmer22}, which has served as a unique experimental fingerprint of nematic fluctuations.

Similarly, the strong coupling of electronic and lattice degrees of freedom makes such electronic systems particularly susceptible to tuning by physical pressure. Driven by novel developments of pressure-cell technology for tuning quantum materials by hydrostatic and uniaxial pressure \cite{Gati20,Hicks14}, important discoveries have been made in the field of quantum materials. Since the application of pressure does not introduce additional disorder into the system, pressure tuning has been instrumental in exploring the fundamental properties of quantum materials, like superconducting $T_c$, in clean systems \cite{Steppke14Sr214}.

As a result, the combination of the two -- the ability to measure elastic properties while tuning the physical pressure -- has proven to be a powerful addition to the toolbox for studying quantum materials. Recently, measurements of the lattice elasticity as a function of pressure have led to important insights into the fundamental question of the role of the lattice in electronic matter \cite{Zacharias15a,Zacharias15b,Sarkar23} and to the identification of possible functionalities \cite{Sypek17}. For example, in correlated quantum materials subjected to pressure tuning, nonlinear elastic behavior has been observed. This experimental observation was taken as a strong indication that the lattice profoundly affects the properties of the electronic system and vice versa. Notable experimental examples in this respect include the observation of the breakdown of Hooke's law \cite{Gati16Hookes,Gati18} around the finite-temperature critical endpoint of the Mott metal-insulator transition in an organic conductor. A huge lattice softening has also been observed\cite{Noad23} at the pressure-induced electronic Lifshitz transition \cite{Sunko19} in Sr$_2$RuO$_4$.

The conclusions above were obtained from measurements of stress-strain relationships under continuously-tuned pressure \cite{Hicks14,Barber19Piezo,Gati20,Agarmani22hydro,Manna12hydro}, e.g., in piezoactuator-driven pressure cells \cite{Hicks14}, performed in the demanding cryogenic environment required for the study of quantum materials. These measurements were made possible by several recent advances in measuring both the applied stress, $\sigma$, and the resulting strain, $\epsilon$, with high precision. In general, stresses and strains are tensor quantities ($\sigma_{ij}$ and $\epsilon_{kl}$) and are related by the compliance matrix $S_{ijkl}$ (or the inverse elastic constant matrix $C_{ijkl}$), i.e., $\sigma_{ij}\,=\,\sum_{k,l} S_{ijkl} \epsilon_{kl}$. When stress, $\sigma$, is applied along a specific direction and the deformation, $\epsilon$, is measured along the same direction, $i\,=\,j\,=\,k\,=\,l$ and the corresponding  modulus $S_{iiii}$ (or $S_{ii}$ in Voigt notation) \cite{Luethi05} is the Young's modulus, which we denote as $E$ throughout this manuscript. It is obtained experimentally from the measured stress-strain relationships via

\begin{equation}
    E = \frac{d \sigma_{ii}}{d \epsilon_{ii}}.
    \label{eq:def-youngsmodulus}
\end{equation}

\noindent $E$ remains unchanged with $\epsilon$ in linear elastic systems, i.e., in systems that obey Hooke's law of elasticity. In contrast, the hallmark of non-linear elastic systems is that $E$ varies with strain \cite{Noad23,Gati16Hookes}.

 \begin{figure}
    \centering
    \includegraphics[width=0.7\columnwidth]{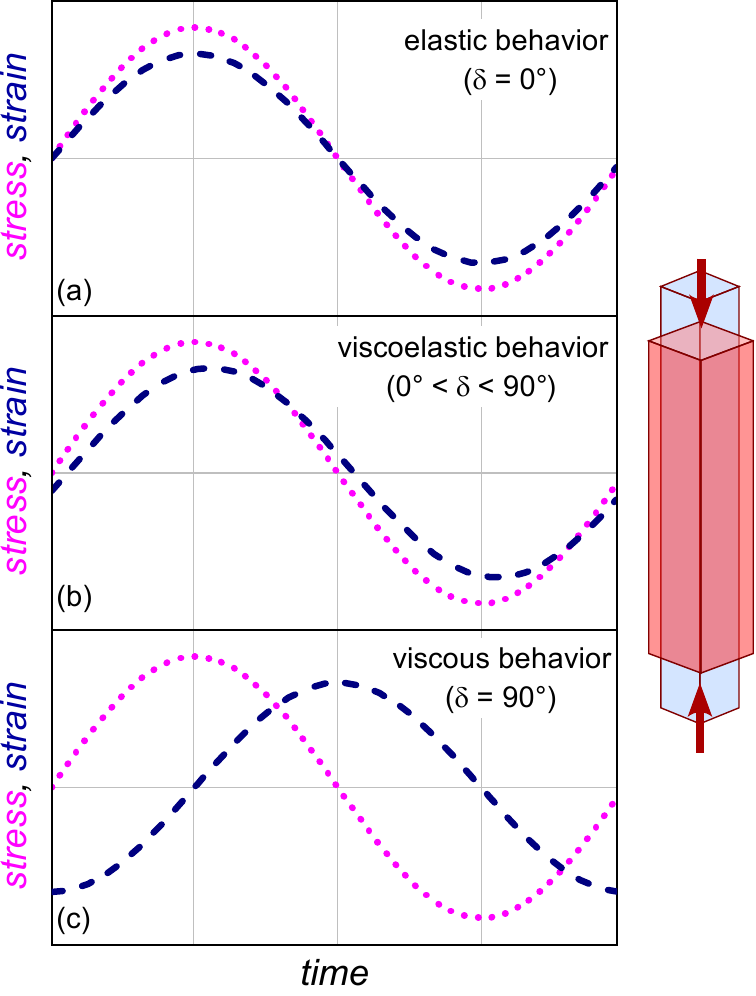}
    \caption{Schematic representation of a dynamic elastic modulus measurement. An a.c. stress (dotted line) is applied to a sample, and the resulting a.c. strain (dashed line) is measured. The phase difference, $\delta$, between the applied stress and the measured strain characterizes the degree of viscous and elastic behavior. For $\delta\,=\,0^\circ$ ($\delta\,=\,90^\circ$), a system exhibits purely elastic (viscous) behavior. If $0\,^\circ<\,\delta\,<\,90\,^\circ$, a material is viscoelastic, i.e., it shows an elastic response similar to that of solids and liquids simultaneously. Note that viscoelastic deformation is still fully recoverable, whereas plastic deformation is characterized by unrecoverable deformation.}
    \label{fig:schematic-viscoelasticity}
\end{figure}

In this paper, we introduce a new method to measure $E(\epsilon)$ in piezoactuator-driven uniaxial pressure cells. In our new approach, we make use of low-frequency a.c. stresses and strains \cite{Ikeda2019ECE,Li22ECE} to determine the a.c. Young's modulus. In fact, at ambient pressure, so-called Dynamical Mechanical Analyzer (DMA) spectroscopy measurements, in which the real and imaginary elastic moduli, $E^\prime$ and $E^{\prime\prime}$, are determined by the application of low-frequency, low-amplitude forces and measurements of the resulting strains (see Fig.\,\ref{fig:schematic-viscoelasticity}), are well-established (see e.g. Ref.\,\citen{Salje11}).  In general, as shown schematically in Fig.\,\ref{fig:schematic-viscoelasticity}, a sinusoidal a.c. stress, $\sigma_{ac} (t)\,=\,\sigma_{ac,0} \sin(\omega_p t)$, induces an a.c. strain response, $\epsilon_{ac} (t)\,=\,\epsilon_{ac,0} \sin(\omega_p t - \delta)$, with a phase lag, $\delta$. $\delta$ can take different values (see Fig.\,\ref{fig:schematic-viscoelasticity}): for purely elastic behavior, an instantaneous strain response is expected ($\delta\,=\,0^\circ$ and  $E^\prime\,\neq\,0, E^{\prime\prime} = 0$), whereas liquids are characterized by a purely viscous behavior with $\delta\,=\,90^\circ$ and $E^\prime\,=\,0, E^{\prime\prime}\neq 0$. When $\delta$ takes values between $0^\circ$ and $90^\circ$, the stress-strain response is classified as viscoelastic and both $E^\prime$ and $E^{\prime\prime}$ are finite. These DMA methods are used intensively in the study of viscoelastic properties \cite{Patra20DMA,Venkategowda22DMA} of soft materials, tissues, biomaterials, or polymers, e.g., to extract characteristic energies of glassy freezing processes of structural entities. In the context of rigid solids, the study of the dynamic moduli, $E^\prime$ and $E^{\prime\prime}$, has mainly been  employed to study ferroelastic phase transitions, where strain acts as the primary order parameter. From the frequency, amplitude and temperature dependence of $E^\prime$ and $E^{\prime\prime}$, the contribution of microstructural changes to the macroscopic elastic Young's modulus has been deduced (e.g., due to domains). In this respect, the observation of superelasticity in ferroelastic materials was a particularly relevant discovery \cite{Kityk00,Schranz08,Schranz21domains}. Here, the stress-induced movement of domain walls \cite{Schranz08} triggers significant length changes and thus results in a large `superelastic' softening.

The new setup described in this paper combines the concepts of DMA measurements with the ability to precisely apply static tuning uniaxial pressures in piezoactuator-driven uniaxial pressure cells \cite{Hicks14,Barber19Piezo}. These cells are nowadays widely used in the field of quantum materials, because they are compatible with a low-temperature and high-magnetic field environment and allow for \textit{in situ} d.c. and a.c. stress tuning \cite{Hicks14,Ikeda2019ECE}. In this context, DMA-type measurements are very promising to investigate, e.g., the character of phase transitions, the role of stress-induced domain dynamics \cite{Schranz97} or other collective effects \cite{Müller11noise,Wang22visco}. In Section \ref{sec:experimentalsetup}, we first describe our experimental setup used for the dynamic Young's modulus measurements and the data analysis procedure. Then, we show in Sec.\,\ref{sec:Proofofprinciple} our proof-of-principle experiments on Sr$_2$RuO$_4$, where we demonstrate that the magnitude of the Young's modulus obtained in our setup is in very good agreement with previous literature results from a d.c. technique \cite{Noad23}. We conclude the paper by discussing the experimental results of a finite imaginary part of the Young's modulus under high uniaxial compression in the magnetic phase of Sr$_2$RuO$_4$ as an illustration of the DMA-type analysis, which is now possible with our setup.

\section{Experimental Setup and Data analysis}
\label{sec:experimentalsetup}

\subsection{The uniaxial pressure cell and determination of the static Young's modulus}

\begin{figure}
    \centering
    \includegraphics[width=.9\columnwidth]{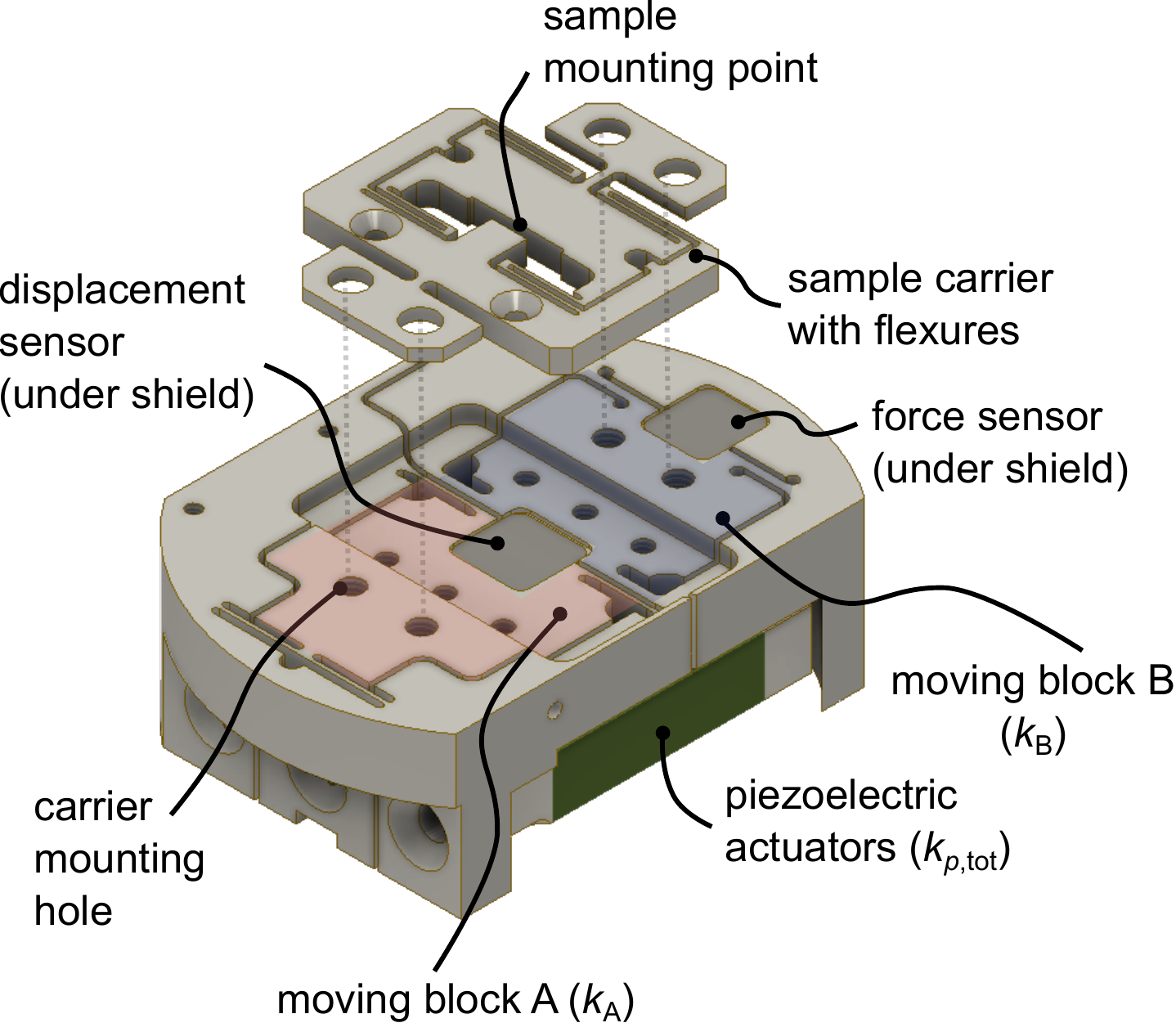}
    \caption{Piezoactuator-controlled uniaxial pressure cell \cite{Barber19Piezo} used to measure the stress-strain relation in quantum materials. In this cell, a set of piezoelectric actuators (green) generate a displacement across a gap between moving block A (highlighted in red) and moving block B (highlighted in blue). The sample carrier, on which the sample is mounted, connects A and B so that the displacement is transmitted to the sample. The displacement of the gap is measured by a capacitive sensor mounted under a shield. In addition, the cell houses a capacitive force sensor, which is used for d.c. stress-strain measurements. The cell's spring constant is determined by that of the moving blocks ($k_A$ and $k_B$) as well that of the actuators ($k_\textrm{P,tot}$).}
    \label{fig:pressure-cell}
\end{figure}

\begin{figure*}
    \centering
    \includegraphics[width=\textwidth]{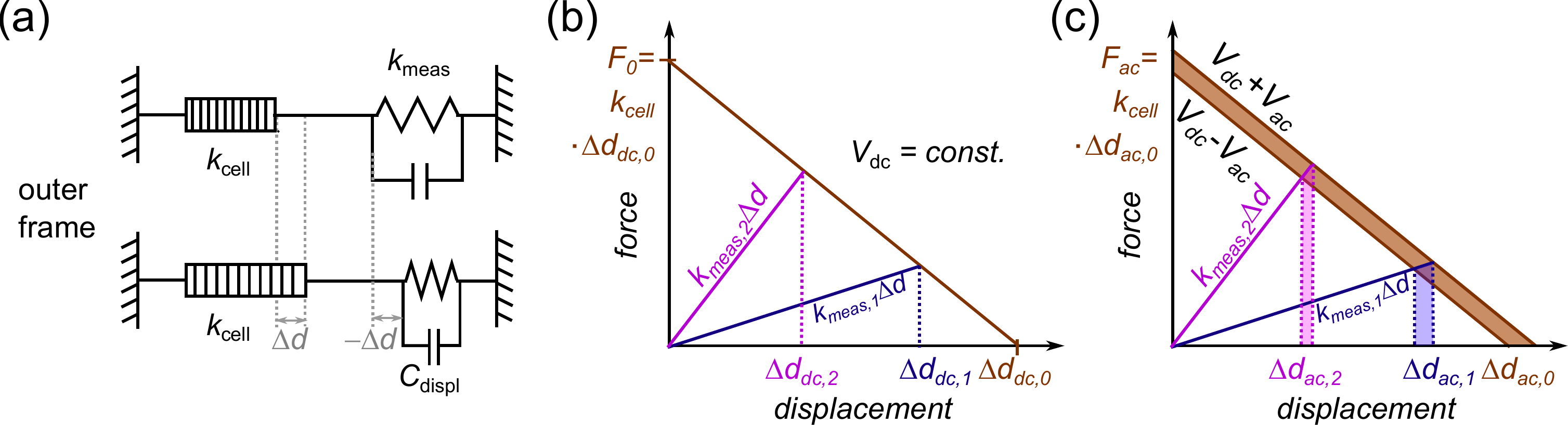}
    \caption{(a) Simplified schematic diagram of the uniaxial pressure cell and its working principle. The cell with spring constant $k_\textrm{cell}$ applies varying forces to the `load' spring $k_\textrm{meas}$, whose magnitude is to be determined in the experiment (see text for details). The displacement of the load spring, $\Delta d$, can be measured by a capacitive displacement sensor and its magnitude is identical to the displacement created by the cell; (b) Working curve of applied force, $F$, vs. generated displacement, $\Delta d_{dc}$, of the piezoactuator-driven uniaxial pressure cell at a constant supply voltage on the piezoelectric actuators, $V_\textrm{dc}$. The working curve is determined by the parameters $k_\textrm{cell}$, the maximum displacement at this $V_{dc}$ without a load, $\Delta d_\textrm{dc,0}$, and the maximum generated force, $F_{0}$, in case of an infinite load. It is shown by the brown line following the form $F\,=\,F_0 -k_\textrm{cell} \Delta d_{dc}$. The magnitude of the load spring constant, $k_\textrm{meas}$, determines the force/displacement provided by the cell at this particular voltage, as seen from the working curve. This is illustrated by showing two linear stress-strain relations, corresponding to two different values of $k_\textrm{meas}$. The intersection of these stress-strain relationships with the working curve determine the applied force and created displacement. The larger $k_\textrm{meas}$ is, the smaller (larger) the created displacement (applied force) will be; (c) Working curve of the piezoactuator-driven uniaxial pressure cell when an a.c. voltage, $V_\textrm{ac}$, is applied in addition to the d.c. voltage, $V_\textrm{dc}$. The working range is now delineated by the two working curves at $V_\textrm{dc}\,\pm\,V_\textrm{ac}$. The alternating $V_\textrm{ac}$ causes an alternating displacement, $\Delta d_\textrm{ac,0}$, created at zero force. The alternating displacement, $\Delta d_\textrm{ac}$, induced by $V_\textrm{ac}$ at finite load spring constant, $k_\textrm{meas}$, is related to the magnitude of $k_\textrm{meas}$, see text for details.}
    \label{fig:setup}
\end{figure*}

We use a uniaxial pressure cell, which is shown schematically in Fig.\,\ref{fig:pressure-cell} and is similar to the one described in Ref.\, \citen{Barber19Piezo}, for an \textit{in-situ} control of the pressure applied to the sample. Here, we briefly review the working principle of the pressure cell.

The cell uses piezoelectric actuators to apply the force to the sample. In the piezocartridge, three piezoelectric actuators are combined to either apply compressive or tensile forces to the sample, depending on the sign of the applied voltage. In the present design, the two outer actuators apply compression to the sample upon application of a positive voltage, whereas the inner one applies tension (see Fig.\,1 of Ref.\,\citen{Barber19Piezo}). Accordingly, we refer to the two actuators (one actuator) as compression (tension) stacks (stack) throughout the manuscript.

The piezoelectric actuators drive the motion of the moving block A (see Fig.\,\ref{fig:pressure-cell}), which is guided by flexures with a low longitudinal spring constant. This changes the size of the gap between block A and B. When a sample is mounted between these two blocks, it is either compressed or stretched. In the present case, we mount our sample on a sample carrier that connects between moving block A and B (details on the sample carrier design and implications for the data analysis will be provided in Sec.\,\ref{sec:dataanalysis}). The purpose of the sample carrier is to facilitate sample exchange (see e.g. Ref.\,\citen{Jerzembeck22caxis}). A capacitive sensor is mounted below the gap between the moving block A and B to measure the relative displacement of moving blocks, which is related to the sample strain.

The cell contains a second capacitive sensor that acts as a force sensor. The moving block B is connected to the frame of the cell by flexures having a larger longitudinal spring constant compared to the flexures of the moving block A (see Appendix \ref{sec:app1} for details). The capacitor measures the displacement of these flexures, which can then be converted into the applied force using the known spring constant of these flexures.

The d.c. capacitances of the displacement and force sensors can be measured separately using high-precision capacitance bridges, such as the model AH2550A from the company Andeen-Hagerling. As described in detail in Ref.\,\citen{Noad23}, these measurements, together with a precise knowledge of the sample dimensions, can be used to calculate the applied stress, $\sigma$, and the resulting strain, $\epsilon$, along the direction of the applied force. Taking the derivative of $\sigma$ with respect to $\epsilon$ in the post-processing analysis then yields the Young's modulus, $E(\epsilon)$, see Eq.\,\ref{eq:def-youngsmodulus}. Since these measurements are based on static measurements of $\sigma$ and $\epsilon$, we will refer to them throughout the text as \textit{static} Young's modulus measurements or, in short, the d.c. method.




\subsection{Working curve of the piezoactuator-driven stress cell}

The working principle of our new a.c. technique is based on the fact that piezoelectric actuators themselves can be used to measure the spring constant of the spring on which they exert force, called the  `load spring'. This capability is rooted in their characteristic working curve \cite{Barber19Piezo}, where key working parameters of piezoelectric actuators, such as the supplied force and the created displacement depend on the spring constant of the load spring.


 In Fig.\,\ref{fig:setup}\,(a), we show the simplified version of our experimental setup to illustrate the concept. The cell with spring constant $k_\textrm{cell}$, which contains the piezoelectric actuators, pushes against a load spring with spring constant  $k_\textrm{meas}$. $k_\textrm{meas}$ is the key quantity that is to be determined in the experiment, which can be converted to the Young's modulus, $E$, of the sample. The working curve of the cell is determined by $k_\textrm{cell}$ and is shown by the brown line in Fig.\,\ref{fig:setup}\,(b). At no load, i.e., $k_\textrm{meas}\,=\,0$, the cell provides a large displacement, $\Delta d_\textrm{dc,0}$, at essentially no force for a given, fixed piezoactuator voltage, $V_\textrm{dc}$. In case of an infinitely stiff sample with $k_\textrm{meas}\,\rightarrow\,\infty$, the cell provides a maximum force of $F_{0}=\,k_\textrm{cell}\Delta d_\textrm{dc,0}$ for the same $V_\textrm{dc}$, but no displacement. Accordingly, the displacement, $\Delta d_\textrm{dc}$, which is generated by the cell and delivered to the load spring, is a function of $k_\textrm{meas}$. For illustration, we also include in Fig.\,\ref{fig:setup}\,(b) two stress-strain curves corresponding to two different load springs, $k_\textrm{meas,1}$ and $k_\textrm{meas,2}\,>\,k_\textrm{meas,1}$. The applied force (displacement) is greater (smaller) for $k_\textrm{meas,2}$ compared to $k_\textrm{meas,1}.$ 

This simplified picture describes the working characteristic of a cell at a fixed piezoactuator voltage for different values of $k_\textrm{meas}$. In our experiment, we want to measure the changes of $k_\textrm{meas}$ of a single sample with changing stress and strain, which is achieved by varying the static voltage on the piezoelectric actuator.

Specifically, we are interested in obtaining the dynamic moduli from an a.c. experiment. To this end, we apply a small a.c. voltage to the piezoelectric actuators, which creates oscillating stresses and strains. Whereas the a.c. component serves to \textit{probe} the Young's modulus, we use the d.c. voltages on the piezoelectric actuators to \textit{tune} a given material. The extended working curve illustrating the situation in our experiment is shown in Fig.\,\,\ref{fig:setup}\,(c). The working range of the cell in the presence of an a.c. voltage, $V_\textrm{ac}$, is now delineated by two parallel lines, corresponding to the working curves at $V_\textrm{dc}\,\pm\,V_\textrm{ac}$ (see brown area). In analogy to the previous discussion, the a.c. displacement is given by $\Delta d_\textrm{ac,0}$, when $k_\textrm{meas}\,=\,0$, and will be zero, when $k_\textrm{meas}\,\rightarrow\,\infty$. It follows that $\Delta d_\textrm{ac}$ is directly related to $k_\textrm{meas}$ for constant amplitude of $V_\textrm{ac}$. Specifically, the larger $k_\textrm{meas}$, the smaller $\Delta d_\textrm{ac}$ will be. Since the a.c. voltage is used to measure $k_\textrm{meas}$, the d.c. voltage can be used independently to tune the elastic properties of the material under investigation.

\subsection{Electronic setup for determining the dynamic Young's modulus using the new a.c. method}
\label{sec:electronics}

\begin{figure*}
    \centering
    \includegraphics[width=0.8\textwidth]{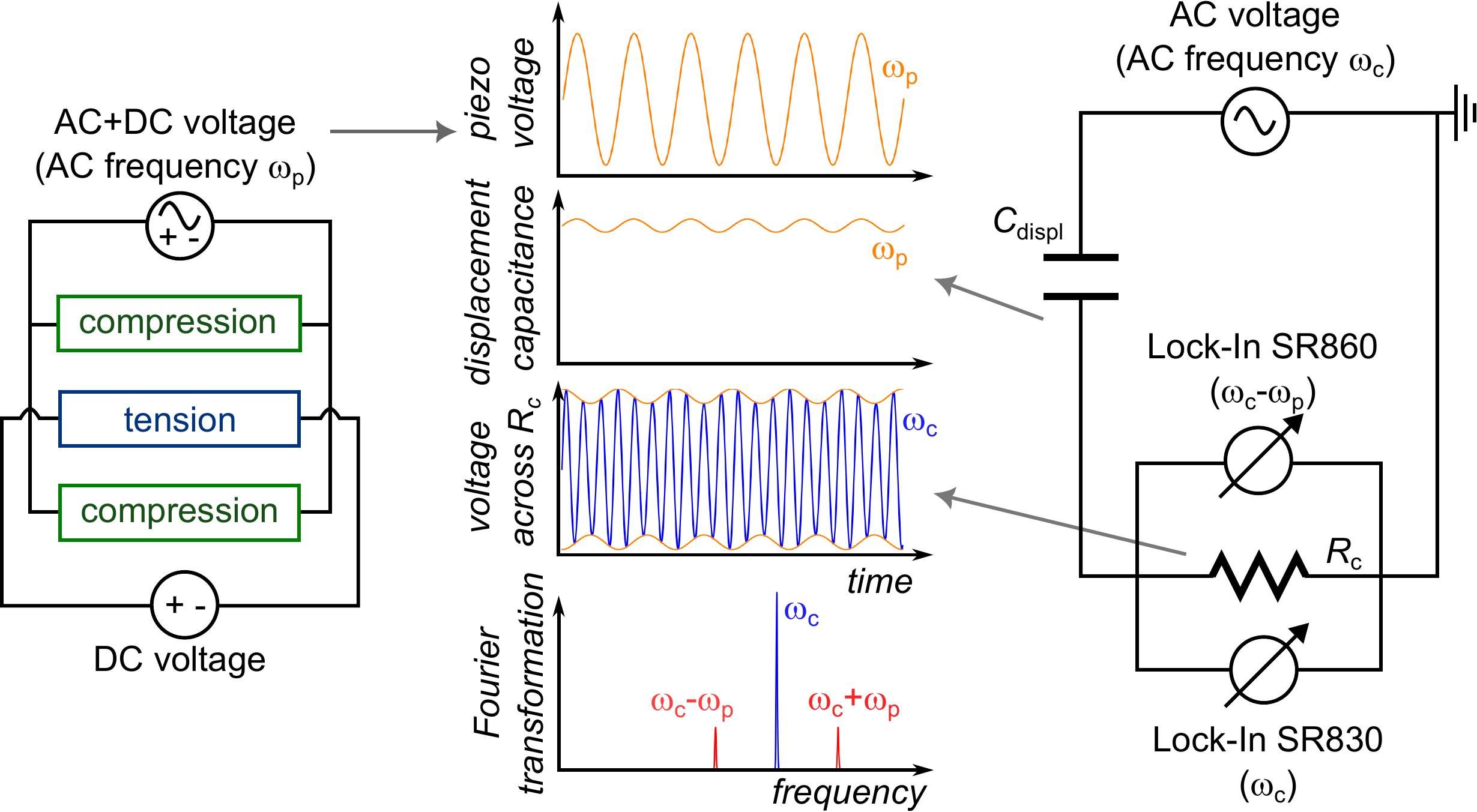}
    \caption{Electronic setup for determining the a.c. displacement, $\Delta d_\textrm{ac}$, which is related to the sample's Young's modulus. The piezocartridge, which drives the uniaxial pressure cell, consists of two compression stacks and one tension stack. A d.c. voltage is applied to all stacks to control the static uniaxial pressure on the sample. The d.c. voltage on the compression stacks is superimposed by a small a.c. voltage with frequency $\omega_p$. The a.c. modulation of the voltages on the compression stack results in an a.c. change of the capacitance of the displacement sensor, $C_\textrm{displ}(t)\,=\,C_{dc}+C_{ac}(t)$, with same frequency $\omega_p$. To measure the contributions of $C_\textrm{dc}$ and $C_\textrm{ac}$ to $C_\textrm{displ}$ independently, a home-built capacitance bridge, consisting of an a.c. voltage supply with frequency $\omega_c$, a resistance $R_C$ and two Lock-in amplifiers (Stanford Research SR860 and SR830), is used. The voltage across the resistor (blue line) corresponds to the modulated signal of signals with frequencies $\omega_p$ and $\omega_c$ (see the orange enveloping curve).  As a result, as shown in the Fourier transformation of the voltage signal, the signal is composed of signals at $\omega_c$ (blue line) as well as at $|\omega_c\,\pm\,\omega_p|$ (red line). The SR830 reads out the component of the voltage signal at $\omega_c$, which is related to the d.c. capacitance value (blue line), whereas the SR860 in Dual Reference mode picks up the signal at the beat frequency $\omega_c-\omega_p$, which is proportional to the a.c. capacitance, see text for details.}
    \label{fig:electronics}
\end{figure*}

Thus, to directly measure $k_\textrm{meas}$, our technique relies on the accurate determination of $\Delta d_\textrm{ac}$, which is a mechanical modulation induced by a finite piezoactuator voltage $V_\textrm{ac}$, through a capacitive measurement. To this end, we have designed a home-built capacitance bridge, which allows $\Delta d_\textrm{ac}$ to be measured simultaneously with $\Delta d_\textrm{dc}$ from a single displacement capacitor. The ability to measure both is important to determine the probing strain (related to $\Delta d_\textrm{ac}$) separately from the tuning strain (related to $\Delta d_\textrm{dc}$).

The detailed working principle of the bridge is shown in Fig.\,\ref{fig:electronics}. A small a.c. voltage is superimposed on the d.c. voltage of one of the stacks. In our case, the voltage on the compression stacks is superimposed by an a.c. modulation with frequency $\omega_{p}$, i.e., $V(t)\,=\,V_\textrm{dc} + V_\textrm{ac}\sin(\omega_{p}t)$. This, in turn, induces a mechanical displacement of the gap, which is measured through a capacitive displacement sensor with capacitance $C_\textrm{displ}$. The distance of the capacitor plates then follows the form $d\,=\,d_{dc} + \Delta d_{ac}\sin(\omega_{p}t)$ with $d_{dc}\,=\,d_{dc,0}+\Delta d_{dc}$ and $d_{dc,0}$ the initial distance of the capacitor plate. Accordingly, the time dependence of $C_\textrm{displ}$ follows as $C_\textrm{displ}\,=\,C_{dc} + C_{ac}\sin(\omega_{p}t)$. In a first approximation, if $\Delta d_\textrm{ac}\,\ll\,d_\textrm{dc}$, these capacitances are related to the displacements by

\begin{equation} 
    \frac{\Delta d_{ac}}{d_{dc}} = \frac{C_{ac}}{C_{dc}}.
    \label{eq:dis_cap_fraction}
\end{equation}

 To extract $C_\textrm{ac}$ and $C_\textrm{dc}$ from $C_\textrm{displ}$, we use a circuit consisting of a voltage source operating at a second frequency, $\omega_c$, and two lock-in amplifiers from Stanford Research Systems: specifically, a SR830 model and a SR860 model (see Fig.\,\ref{fig:electronics}). In general, applying an a.c. voltage with frequency $\omega_{c}$ to a capacitor with time-independent value $C_\textrm{dc}$ generates an a.c. current with $\omega_c$. In our present case, where $C_\textrm{displ}$ is time-dependent, the returned current, with characteristic frequency $\omega_c$, will be further modulated by $\omega_p$. This modulated current is passed through a resistor with $R_c$, where it produces a voltage, $V_R$, which reads as

\begin{equation}
\begin{split}
        V_R &= C_{dc} V_0 R_c \omega_c \cos{(\omega_c t)} + C_{ac} V_0 R_c \\ & \times [\omega_c \cos{(\omega_c t)} \sin{(\omega_p t)} + \omega_p \cos{(\omega_p t)} \sin{(\omega_c t)}].
\end{split}
    \label{eq:Voltage_over_resistor}
\end{equation}

The main task of the two lock-in setup is to perform an electrical demodulation of the signal \cite{Hristov18demod}. The first term is directly proportional to $C_\textrm{dc}$ and thus to $d_\textrm{dc}$. It corresponds to the high-frequency signal represented by the blue curve in Fig.\,\ref{fig:electronics} (for $\omega_c\gg\omega_p$). We read out this component of the voltage in Eq. \ref{eq:Voltage_over_resistor} by locking the SR830 to $\omega_{c}$, such that $V_\textrm{SR830}\,=\,C_{dc}V_0R_c\omega_c$. The second and third terms in Eq. \ref{eq:Voltage_over_resistor} are products of waves with the characteristic side-band frequencies $|\omega_c \pm \omega_p|$ (see also red line in the Fourier transform in Fig.\,\ref{fig:electronics}). To read out these voltages, we use the SR860 in Dual Reference mode, which measures the voltage at the frequency $|\omega_c-\omega_p|$. It then follows that

\begin{equation} 
    V_\textrm{SR860} = \frac{ C_{ac} V_0 R_c |\omega_c - \omega_p|}{2},
    \label{eq:acVoltage}
\end{equation}

 \noindent which can be directly converted to $C_\textrm{ac}$. The signal-to-noise ratio of $V_\textrm{SR860}$ will be larger, the greater the difference between $\omega_p$ and $\omega_c$.

Using the known calibration of the displacement sensor, i.e., the functional form of $C_\textrm{displ}$ vs. $d$, $\Delta d_\textrm{ac}$ and $\Delta d_\textrm{dc}$ can now be calculated from the measured voltages $V_\textrm{SR830}$ and $V_\textrm{SR860}$. In Fig.\,\ref{fig:spring}\,(a), we show an example curve of $\Delta d_\textrm{ac}$ vs. tuning displacement, $\Delta d_\textrm{dc}$, taken on Sr$_2$RuO$_4$. We will discuss the implications of the data in detail below in Sec.\,\ref{sec:Proofofprinciple}. For now, the plot clearly shows that $\Delta d_\textrm{ac}$ is not constant when $\Delta d_\textrm{dc}$ is changed, and that the changes in $\Delta d_\textrm{ac}$ can be clearly resolved in our setup with a resolution of $\approx\,0.2\,$nm. As explained above, the changing $\Delta d_\textrm{ac}$ reflects the changing $k_\textrm{meas}$ (or in other words, the changing Young's modulus, $E$, of the sample).

As an aside, we would like to remark that the precise determination of $\Delta d_\textrm{ac}$ and hence $\epsilon_\textrm{ac}$, which we demonstrate here, is also crucial for quantitative measurements of the elastocaloric effect \cite{Ikeda2019ECE,Li22ECE}, where the temperature change, $\Delta T$, induced by a finite $\epsilon_\textrm{ac}$, is measured.

\subsection{Data Analysis}
\label{sec:dataanalysis}

We now discuss how to convert the measured $\Delta d_\textrm{ac}(\Delta d_\textrm{dc})$ into absolute values of the Young's modulus $E(\epsilon)$. This requires two main steps: (i) converting $\Delta d_\textrm{ac}$ to the measured spring constant, $k_\textrm{meas}$, and (ii) extracting the sample's spring constant from $k_\textrm{meas}$. This second step is necessary because $k_\textrm{meas}$ contains contributions from, e.g., the mounting glue and the sample carrier, in addition to the contribution of the sample. This part of the analysis is also required when analyzing the data from the d.c. method and is described in detail in the Supplementary Information of Ref.\,\citen{Noad23}.

\begin{figure}[h!]
    \centering
    \includegraphics[width=.8\columnwidth]{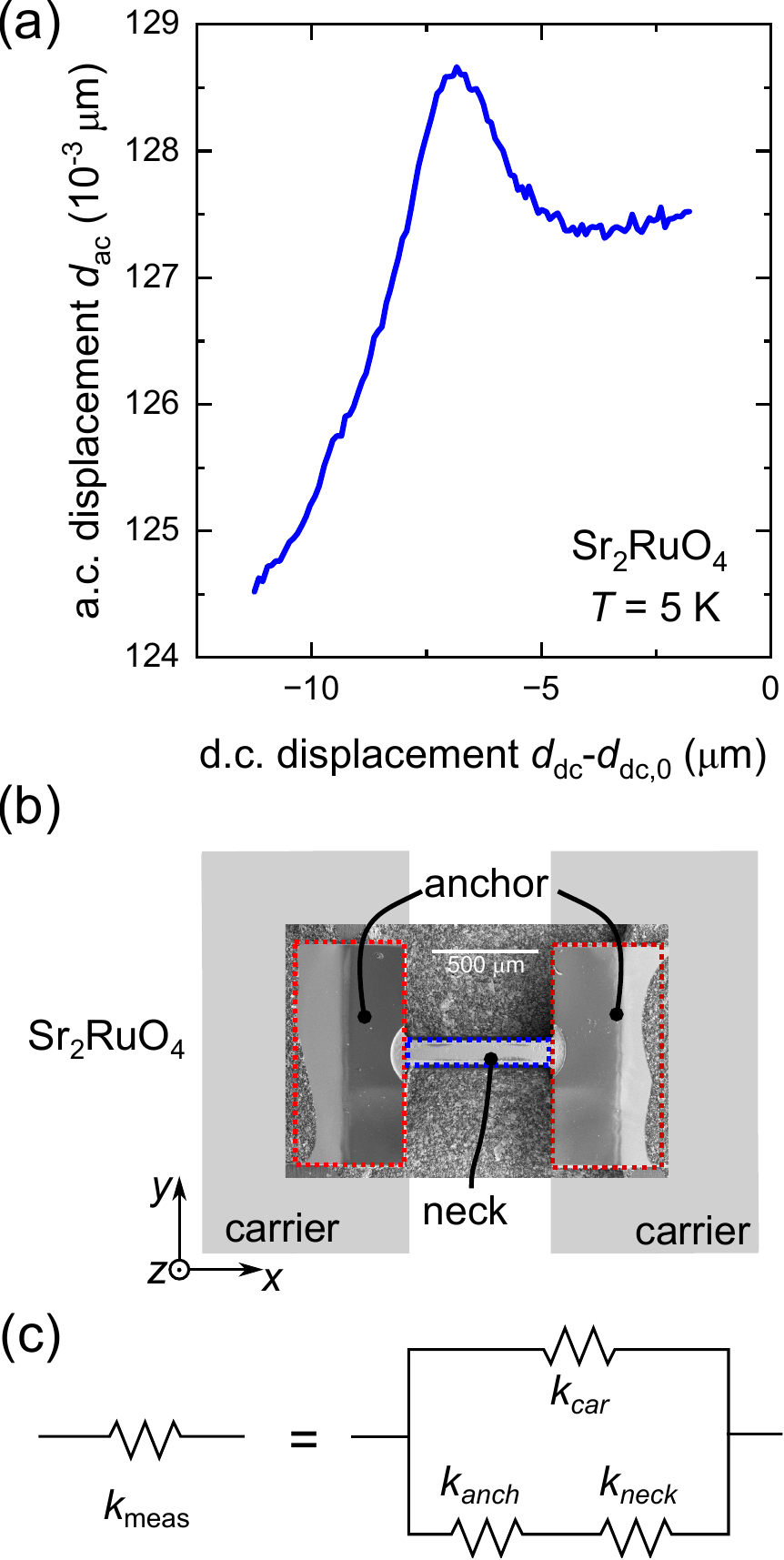}
    \caption{
    (a) Experimental data of the a.c. displacement, $\Delta d_\textrm{ac}$, as a function of the d.c. tuning displacement, $\Delta d_\textrm{dc}$, induced by a the application of a d.c. and an a.c. voltage on the piezoelectric actuators. $d_{ac}$ and $d_{dc}$ were obtained by reading out the displacement capacitive sensor in the piezoactuator-driven pressure cell with our home-built capacitance bridge. This example dataset was recorded on Sr$_2$RuO$_4$ at a temperature of $T\,=\,$ 5\,K, a frequency $f_p$ of 167\,Hz and an a.c. amplitude $V_{ac}$ of 5\,V;
    (b) Preparation and mounting of samples for stress-strain measurements in uniaxial pressure cells. The sample (in the present case: Sr$_2$RuO$_4$) is cut into a narrow neck with wide anchor tabs using a Plasma Focused Ion Beam. The anchors are epoxied to the sample carrier. The sample shape ensures a rapid crossover from the low-stress region in the anchor tabs to the high-stress region in the neck and minimizes the stress in the mounting epoxy;
    (c) Due to the shape of the sample, the sample's spring constant can be modeled to a good approximation as a set of two discrete springs in series, i.e., the anchor (spring constant $k_\textrm{anch}$) and the neck spring (spring constant $k_\textrm{neck}$). In addition, the small carrier spring constant, $k_\textrm{car}$, contributes to the measured spring constant, $k_\textrm{meas}$, and is in parallel to $k_\textrm{anch}$ and $k_\textrm{neck}$.}
    \label{fig:spring}
\end{figure}

For the first step, we refer back to the working curve of the cell, which has been introduced in Sec.\,\ref{sec:electronics} and is shown in Fig.\,\ref{fig:setup}\,(c). For example, we can compare the a.c. displacement in case of an empty cell, $\Delta d_\textrm{ac,0}$, with the displacement, $\Delta d_\textrm{ac}$, when the cell works against a load spring with $k_\textrm{meas}$. Clearly, $\Delta d_\textrm{ac}\,<\,\Delta d_\textrm{ac,0}$ and it can be deduced that the ratio determines $k_\textrm{meas}$ via

\begin{equation}
    k_\textrm{meas} = k_\textrm{cell} \left( \frac{\Delta d_\textrm{ac,0}}{\Delta d_\textrm{ac}} -1 \right).
    \label{eq:sp_constant_meas}
\end{equation}

Thus, $k_\textrm{meas}$ can be calculated for all values of $\Delta d_\textrm{ac}$, when $\Delta d_\textrm{ac,0}$ is measured, e.g. when the sample is broken after an experiment. Alternatively, any reference point with finite $k_\textrm{meas}$ can be chosen, as long as the value of $k_\textrm{meas}$ at a particular strain is well known.



The key parameter for converting $\Delta d_\textrm{ac}$ to $k_\textrm{meas}$ is the value of the cell spring constant, $k_\textrm{cell}$, which cannot be determined during the experiment itself. In general, it can be determined from calibration experiments and from simulations, as shown in Ref.\,\citen{Barber19Piezo} for a cell of similar design. Following the same procedure, we determined the cell spring constant of the present cell to be $k_\textrm{cell}\,=\,(3.4\,\pm\,0.5)$\,N/$\mu$m (see Sec.\,\ref{sec:app1} for details). Since titanium, the material of the cell, becomes stiffer upon cooling and calibration experiments were performed at room temperature, we applied a correction factor of 15\% to $k_\textrm{cell}$, when analyzing low-temperature data \cite{Barber19Piezo}. The exact value of $k_\textrm{cell}$ will likely vary from cell to cell, even if the design is technically the same. Once $k_\textrm{cell}$ is determined for a specific cell, e.g., by calibration measurements, it is not expected to change from experiment to experiment and is thus not a free parameter in the analysis. Equation \ref{eq:sp_constant_meas} also shows that the ability to resolve changes in $k_\textrm{meas}$ also depends on $k_\textrm{cell}$. As expected from the working curve, the setup works best when $k_\textrm{cell}$ and $k_\textrm{meas}$ are of similar magnitude.

The second step in obtaining $E(\epsilon)$ from the a.c. data is to extract the spring constant of the sample from $k_\textrm{meas}$. Following the protocol established in Ref.\,\citen{Noad23}, the samples are cut into a narrow neck with wide anchor tabs using a Xe Plasma Focused Ion Beam (see Fig.\,\ref{fig:spring}\,(b)). The necking creates a rapid crossover between regions of low stress in the anchors and high stress in the neck. The sample can therefore be approximated by two springs in series, which we label as $k_\textrm{neck}$ and $k_\textrm{anch}$. We also include the spring constant of the mounting glue in $k_\textrm{anch}$. In the two-spring approximation, $k_\textrm{anch}$ can be calculated from the total measured signal $k_{meas}$ around a reference strain and an independently measured value of $E$ at that reference strain \cite{Noad23}. In the case of Sr$_2$RuO$_4$ over the range of displacements considered here, Ref.\,\citen{Noad23} shows that $k_\textrm{anch}$ can indeed be taken as stress-independent to a good approximation.

In addition, the sample carrier has flexures with spring constants $k_\textrm{car}$, which are in series with $k_\textrm{neck}$ and $k_\textrm{anch}$, as shown schematically in Fig.\,\ref{fig:spring}\,(c). It follows that 

\begin{equation}
    k_\textrm{neck}(\epsilon) = \left[\frac{1}{k_\textrm{meas}(\epsilon) - k_\textrm{car}} -  \frac{1}{k_\textrm{anch}}\right]^{-1}.
    \label{eq:sp_constant_neck}
\end{equation}

\noindent The spring constant of the carrier, $k_{car}\,\approx\,0.02$\,N/$\mu$m, is per design much smaller than typical values of $k_\textrm{meas}$ and can also be calibrated experimentally.

Finally, the Young's modulus is obtained from exact knowledge of the length of the necked region, $l_\textrm{neck}$, and the cross-sectional area of the sample, $A_\textrm{cs}$, via $E\,=\, k_\textrm{neck} l_\textrm{neck}/A_\textrm{cs}$. The strain in the neck, $\epsilon_{xx}$, follows from the measured $d_{dc}$ values via $(k_{\text{meas}}/k_{\text{neck}})\,\times\,({\Delta}d_{dc}/l_{\text{neck}})$. The sample of Sr$_2$RuO$_4$ used for our proof-of-principle measurements shown in Fig.\,\ref{fig:spring}\,(b) was cut to dimensions of $A_\textrm{cs}\,=\,$102\,$\mu$m $\times$ 120\,$\mu$m and $l_\textrm{neck}\,=\,$717\,$\mu$m, with the long edge of the neck oriented along the x\,=\,[1\,0\,0] direction of the crystal, so that $k_\textrm{neck}\,(\epsilon_{xx}\,=\,0)\,=\,$2.72\,N/$\mu$m. The anchor spring constant in our experiments was $k_\textrm{anch}\,=\,$2.36\,N/$\mu$m.

\subsection{Working Parameters of the a.c. Young's modulus setup}

In the following, we specify the working range of the a.c. Young's modulus setup, including accessible frequency, amplitude and temperature range. 

The frequency range of operation is determined by the choice of piezoelectric actuators and the mechanical resonances of the cell. The multi-layer ceramic actuators \cite{PICMAactuators} typically used in the pressure cells operate well in the low-frequency range 1\,Hz\,$\lesssim\,f_p\,=\,\omega_p/(2\pi)\,\lesssim\,$1\,kHz. Future technical developments might be able to extend the frequency range significantly, e.g., by choosing different type of actuators.

The second relevant frequency is that of the capacitance bridge, $f_c$. All measurements in this manuscript were taken with $f_c\,=\,\omega_c/(2 \pi)\,=\,2.297\,$kHz, which is of the same magnitude as the frequency of 1\,kHz used in the Andeen-Hagerling AH2550A capacitance bridges, employed in the earlier d.c. work. 

In terms of the amplitude, the maximum voltage that can be applied to the actuators is temperature dependent. At lowest temperatures, the voltage range can typically be extended\cite{Barber19Piezo} to -300\,V to 400\,V. With such large d.c. voltages, tuning strains of $\,\pm\,$1-2\% may be achieved for a sample of the previously-mentioned dimensions. Since the a.c. voltage produces the probing strain, its magnitude must be chosen to be much smaller than the d.c. voltages. In our proof-of-principle studies on Sr$_2$RuO$_4$, we used a.c. voltages, $V_{ac}$, up to 5\,V at lowest temperatures, corresponding to typical values of stress and strain amplitude of $\sigma_{ac}\,\approx\,10^{-2}\,$GPa and $\epsilon_{ac}\,\approx\,1\,\times\,10^{-4}\,$. $V_{ac}$ and $f_{ac}$ were chosen to optimize signal-to-noise ratio, while ensuring that phase transition features were not significantly smeared out by a large $V_{ac}$.

The piezoacutator-driven uniaxial pressure cells are designed to operate down to very low temperatures, even down to dilution-fridge temperatures. Typically, the lowest-accessible temperature is limited by the cooling power of the fridge and by the heating created by the piezoelectric actuators. In the a.c. setup, heating of the actuators becomes a serious problem at higher frequencies and/or higher a.c. voltage amplitudes. Thus, in practical terms, the lowest accessible temperature may be a trade-off with the frequency/amplitude range of interest.

The a.c. Young's modulus measurements can also be performed in a finite magnetic field, which then allows the simultaneous study of the effect of different tuning parameters on the elastic properties of quantum materials.

\section{Proof-of-Principle Results}
\label{sec:Proofofprinciple}

To demonstrate the functionality of our new a.c. Young's modulus setup, we performed proof-of-principle experiments on the ruthenate Sr$_2$RuO$_4$, whose Young's modulus under finite strain has recently been documented with high precision by d.c. stress-strain measurements \cite{Noad23}. Before discussing our proof-of-principle a.c. data taken on Sr$_2$RuO$_4$, we first present the main aspects of the phase diagram of Sr$_2$RuO$_4$ under uniaxial pressure that are relevant for the present work.


The unconventional superconductor Sr$_2$RuO$_4$ has been extensively studied in uniaxial pressure experiments \cite{Steppke14Sr214, Sunko19, Noad23, Jerzembeck22caxis, Grinenko21uSR, Li22ECE} in the last decade. These studies have uncovered a rich phase diagram under uniaxial stress, $\sigma_{xx}$, applied along the [1\,0\,0] axis of the tetragonal lattice. Upon increasing compression, the Fermi surface of Sr$_2$RuO$_4$, consisting of so-called $\alpha$, $\beta$ and $\gamma$ sheets (see sketches on top of Fig.\,\ref{fig:YMReal}), becomes distorted. The $\gamma$ sheet shows the strongest changes. When a compressive [1\,0\,0] strain of $\epsilon_{xx}\,\approx\,-0.45\%$ is applied, the $\gamma$ sheet undergoes a Lifshitz transition, at which the Fermi surface drastically changes its topology \cite{Sunko19} from a closed to an open configuration. Although the Lifshitz transition involves only a fraction of the conduction electrons, it has recently been shown by Noad \textit{et al.} \cite{Noad23} that these conduction electrons drive a very large lattice softening. The significant renormalization of the Young's modulus upon crossing the Lifshitz transition, that is reported with high accuracy, provides an excellent testbed to benchmark our new a.c. method. In addition, a recent set of experiments \cite{Grinenko21uSR, Li22ECE}, including thermodynamic measurements of the elastocaloric effect, suggests that Sr$_2$RuO$_4$ undergoes a transition into a magnetically-ordered state for higher compression beyond the Lifshitz strain and for temperatures $T\,\lesssim\,$8\,K. This additional phase transition is expected to lead to anomalous behavior in $E(\epsilon_{xx})$.

In the following, we first discuss the results of the magnitude of the dynamic Young's modulus of Sr$_2$RuO$_4$. For low-enough frequencies, it may be expected that the magnitude of the a.c. Young's modulus agrees with the one inferred from d.c. measurements. Afterwards, we discuss our measurements of the phase shift in the dynamic signal as we tune the material into its magnetically ordered regime.

\subsection{Results: Magnitude of the dynamic Young's modulus of Sr$_2$RuO$_4$, determined from the a.c. technique}

\begin{figure}
   \centering
   \includegraphics[width=\columnwidth]{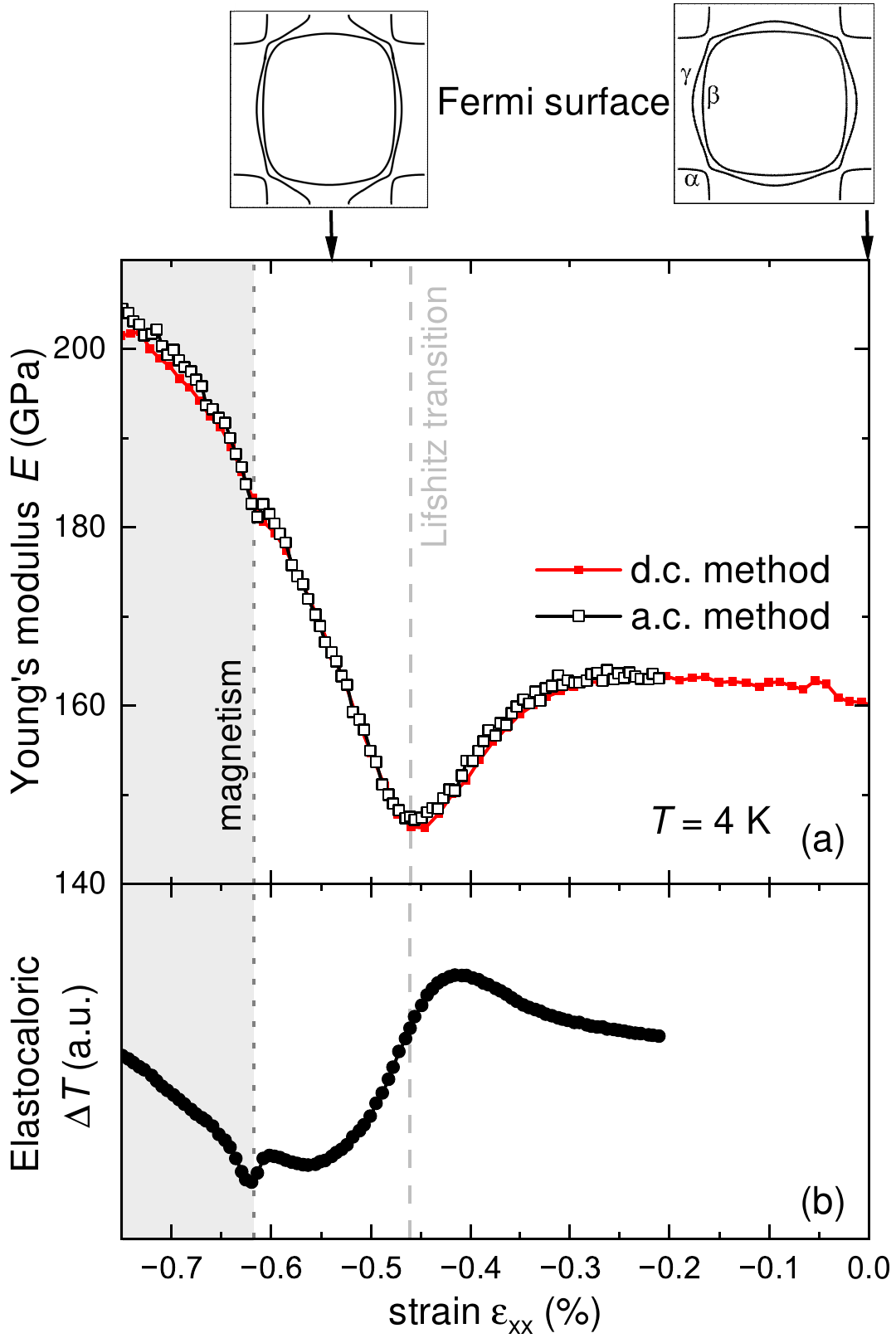}
    \caption{(a) Young's Modulus, $E$, of Sr$_2$RuO$_4$ as a function of strain, $\epsilon_{xx}$, at a temperature of $T\,=\,4$\,K, extracted from the new a.c. method (open symbols) and compared to results from the d.c. method (closed symbols) on the same sample. The data in the a.c. method was taken at a frequency of $f_p\,=\,$167\,Hz and at a piezoactuator a.c. voltage of $V_{ac}\,=\,5\,$V; (b) Elastocaloric temperature oscillation amplitude, $\Delta T$, measured in the same experiment as the Young's modulus data shown in (a). In both panels, the dashed line indicates the strain at which Sr$_2$RuO$_4$ undergoes a Lifshitz transition of the Fermi surface, which is schematically shown in the insets. The dotted line indicates the strain at which Sr$_2$RuO$_4$ undergoes a transition to magnetic order.}
   \label{fig:YMReal}
\end{figure}

In Fig.\,\ref{fig:YMReal}\,(a) we compare the results of Young's modulus measurements conducted at a temperature of 4\,K with the d.c. method \cite{Noad23} (closed symbols) vs. the new a.c. method (open symbols), taken on the exact same sample of Sr$_2$RuO$_4$ in the same experimental run. Clearly, the two data sets as a function of tuning strain, $\epsilon_{xx}$, are in very good agreement. The quantitative agreement is achieved by using $k_\textrm{cell}\,=\,(3.7\,\times\,1.15)\,$N/$\mu$m in the analysis (see Sec.\,\ref{sec:dataanalysis}). This value is within the error bars of the simulation results for $k_\textrm{cell}$ (see Sec.\,\ref{sec:app1}).

Importantly, the data show a set of anomalies associated with the rich phase diagram of Sr$_2$RuO$_4$ under [1\,0\,0] stress. The pronounced softening of $E$ at the electronic Lifshitz transition at around $-0.45\%$ strain is clearly resolved. Upon further increasing compression, a second, albeit smaller, anomaly can be discerned in the $E(\epsilon_{xx})$ data around $\epsilon_{xx}\,\approx\,-0.64\%$. Although the anomaly is visible in both d.c. and a.c. data sets, it is slightly more evident in the a.c. data. That this small drop in $E(\epsilon_{xx})$ does indeed correspond to a thermodynamic phase transition, becomes clear when comparing to the results of elastocaloric measurements, which were performed simultaneously to the $E(\epsilon_{xx})$ measurements. The elastocaloric temperature amplitude, $\Delta T$, manifests an anomaly at the same strain as where the small drop in $E$ occurs. In the previous work by Li \textit{et al.}, this feature was associated with the transition into the magnetically ordered phase \cite{Grinenko21uSR} of Sr$_2$RuO$_4$.

\begin{figure}
    \centering
   \includegraphics[width=\columnwidth]{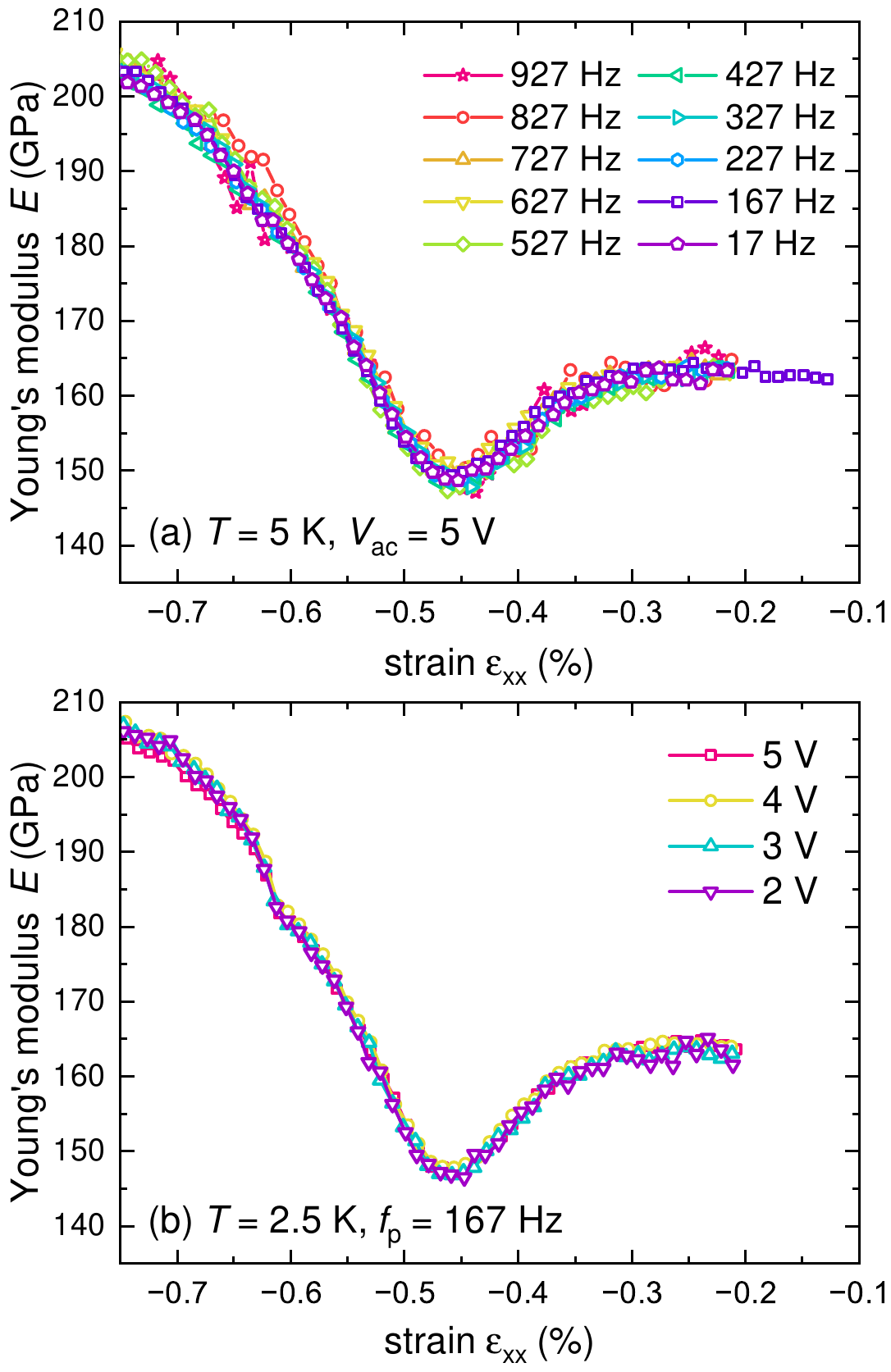}
   \caption{Dependence of the Young's modulus, $E$, of Sr$_2$RuO$_4$ on frequency (a) and stress amplitude (b), as obtained by the new a.c. method. The data are plotted as a function of strain, $\epsilon_{xx}$. In (a), different actuator frequencies, $f_p$, are applied in the range between 17\,Hz and 927 Hz at a temperature of 5\,K and an a.c. voltage of 5\,V. This voltage induces a stress amplitude of $\sigma_{ac}\,=\,$0.027 GPa. In (b), the data for different a.c. voltage amplitudes, $V_{ac}$, between 2\,V and 5\,V are shown at a temperature of 2\,K and an actuator frequency of 167\,Hz. The corresponding stress amplitudes, $\sigma_{ac}$, vary between 0.011 GPa and 0.027 GPa. All data shown were taken with a capacitance-bridge frequency of $f_c\,=\,2.297\,$kHz.}
    \label{fig:YMFreq}
\end{figure}

The a.c. Young's modulus data in Fig.\,\ref{fig:YMReal} were taken at $f_p\,=\,167\,$Hz, $f_c\,=\,2.297\,$kHz, and $V_{ac}\,=\,5\,$V. In Fig.\,\ref{fig:YMFreq}, we now show our data at different $f_p$ (a) and different $V_{ac}$ (b) to demonstrate experimentally that our setup is operational over wider ranges of $f_p$ and $V_{ac}$. Both data sets demonstrate that the magnitude of Young's modulus is essentially independent of the exact frequency or amplitude within the ranges of 17\,Hz\,$\leq\,f_p\,\leq\,927$\,Hz and  2\,V$\,\leq\,V_{ac}\,\leq\,$5\,V. Specifically, in Fig.\,\ref{fig:YMFreq}\,(a), the feature of the Lifshitz transition, which is the prominent feature at $T\,=\,5\,$K, is clearly visible in all data sets at different $f_p$ and the absolute value of $E$ agrees between the data sets within the signal-to-noise ratio. However, the latter is smaller at higher frequency, as expected, since the absolute signal depends on $\omega_c-\omega_p$, as seen in Eq.\,\ref{eq:acVoltage}. 

 We now demonstrate that the chosen $V_{ac}$ are small enough to not smear out phase transitions. To this end, we show in Fig.\,\ref{fig:YMFreq}\,(b) the amplitude dependence of the Young's modulus at a temperature of 2.5\,K, where the feature associated with the magnetic phase at $\epsilon_{xx}\,\sim\,-0.62\,\%$ is more pronounced, and thus serves as a good benchmark for this analysis. For the range of $V_{ac}$ between 2\,V and 5\,V, corresponding to a range of $\sigma_{ac}$ ($\epsilon_{ac}$) between 0.011\,GPa and 0.027\,GPa ($6\,\times\,10^{-5}$ and $1.5\,\times\,10^{-4}$), no significant smearing of Young's modulus anomalies can be detected. Overall, all the data agree quantitatively very well, and only the signal-to-noise ratio changes with decreasing $V_{ac}$, as expected.

\subsection{Results: Phase of the dynamic Young's modulus of Sr$_2$RuO$_4$}

\begin{figure}
    \centering
   \includegraphics[width=\columnwidth]{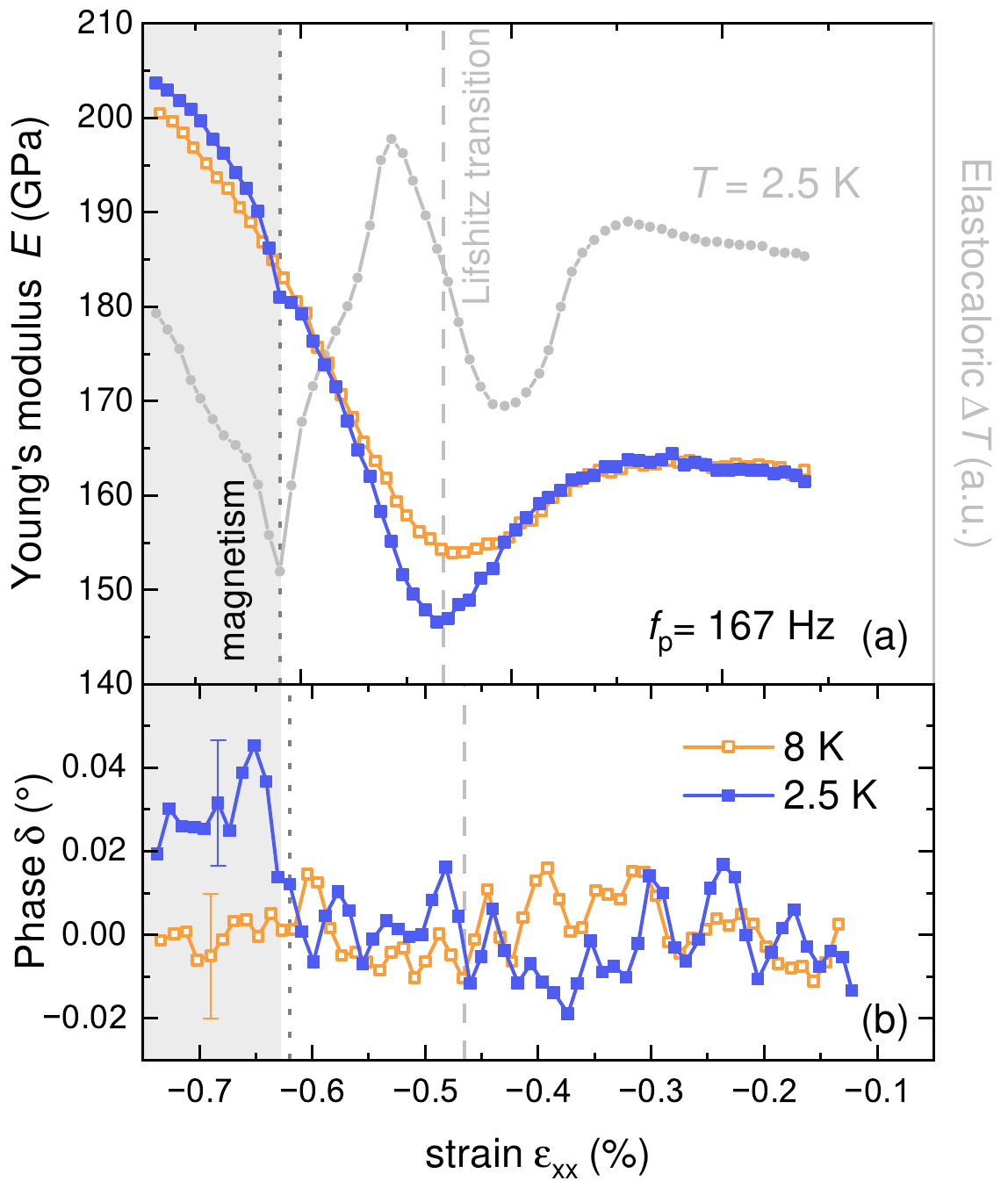}
   \caption{Magnitude (a) and phase (b) of the dynamic Young's modulus of Sr$_2$RuO$_4$  under in-plane strain, $\epsilon_{xx}$. Data of the dynamic modulus are shown at temperatures of 2.5\,K and 8\,K. The strain region, in which Sr$_2$RuO$_4$ is magnetically ordered at 2.5\,K, is indicated by the grey area. The transition into the magnetically-ordered phase at 2.5\,K and $\epsilon_{xx}\,\approx\,-0.62\%$ is also clearly identified in the elastocaloric temperature amplitude, $\Delta T$, which is also included in (a) on the right axis. At 8\,K, Sr$_2$RuO$_4$ remains non-magnetic over the strain range shown here \cite{Li22ECE}. The dashed line indicates the position of the Lifshitz transition at 2.5\,K. All data shown were taken at a frequency of $f_p\,=\,$167\,Hz and an amplitude of $V_{ac}=$\,5\,V. Error bars for the phase value are exemplarily indicated in (b) at strain of $\sim\,-0.7\%$.}
    \label{fig:YMphase}
\end{figure}

We now turn to the additional phase information that is provided by performing our measurements in the a.c. mode. As introduced in Sec.\,\ref{sec:intro}, dynamic measurements reveal information on the real and imaginary part of the Young's modulus ($E^\prime$ and $E^{\prime\prime}$). In particular the latter is of interest to investigate dissipative processes. In the following, we describe our results in this respect by presenting the Young's modulus in terms of its magnitude and phase. Phase and magnitude are related to the real and imaginary part of the modulus via $E^\prime\,=\,\frac{|\sigma_{ac,0}|}{|\epsilon_{ac,0}|} \cos{\delta}$ and $E^{\prime\prime}\,=\,\frac{|\sigma_{ac,0}|}{|\epsilon_{ac,0}|} \sin{\delta}$, i.e., a finite $\delta$ corresponds to a finite $E^{\prime\prime}$. 

Figure \,\ref{fig:YMphase} summarizes our measurement results of $\delta$ in our experiments on Sr$_2$RuO$_4$ at a probing frequency of $f_p\,=\,167$\,Hz and $V_{ac}\,=\,5\,$V. To this end, we now compare the results at two different temperatures, $T\,=\,2.5\,$K and $T\,=\,$8\,K. Whereas Sr$_2$RuO$_4$ was reported to enter a magnetically-ordered phase at high compression at $T\,=\,$2.5\,K, it remains non-magnetic at 8\,K \cite{Grinenko21uSR,Li22ECE}. This is fully consistent with our thermodynamic data, shown in Fig.\,\ref{fig:YMphase}\,(a). At low compression, the magnitude of $E$ as a function of $\epsilon_{xx}$ shows the softening at the Lifshitz transition, and the softening is more pronounced at low temperatures, consistent with the earlier reports \cite{Noad23}. For higher compression, the data taken at 2.5\,K reveal an additional feature at $\epsilon_{xx}\,\approx\,-0.61\%$, associated with the transition into the magnetic state. In contrast, the data set taken at 8\,K shows no features of additional phase transitions besides the Lifshitz transition. The grey area in Fig.\,\ref{fig:YMphase} marks the region of magnetic order at 2.5\,K, as determined by the simultaneous measurements of $E$ and the elastocaloric effect (see grey line, right axis).

In Fig.\,\ref{fig:YMphase}\,(b), we show the behavior of the phase, $\delta$, as a function of $\epsilon_{xx}$ for the same temperatures and at the same frequency. At 8\,K, no change of $\delta$ with strain is observed within the signal-to-noise level over the entire strain range. In contrast, at 2.5\,K, an increase of $\delta$ is observed at the strain, where Sr$_2$RuO$_4$ undergoes the transition into the magnetically-ordered state. Although the changes of $\delta$ are very small, they exceed the signal to noise by more than a factor of two. 

The result of a finite phase between applied stress and resulting strain in the magnetic phase of Sr$_2$RuO$_4$ is a key new insight accessible by extending stress-strain measurements to finite frequencies. It implies that there is an energy dissipation during the (un-)loading stress cycle. Since this phenomenon occurs at low frequencies, compared to typical spin relaxation times, it is likely to be related to interactions between the magnetic order and the crystal lattice, such as domain walls or other collective effects. 
In elemental chromium (Cr), for example, it was found that the pressure-dependent spin-density wave vector shows a certain degree of irreversibility between increasing and decreasing pressure \cite{Ruesnik80Chromium}.
This observation was interpreted in terms of crystal-lattice domain-wall distortions \cite{Fenton80domainsSDW}, which lock the wave vector for small distortions. A similar mechanism may be at work here in Sr$_2$RuO$_4$. Further knowledge of the ordering vector and its strain dependence \cite{Kim23SDW} in Sr$_2$RuO$_4$ will be crucial for understanding the low-frequency response of its elastic constants.

\section{Conclusion and Outlook}

In this work, we described a new experimental technique to determine the dynamic Young's modulus as a function of pressure, frequency and temperature in piezoactuator-driven uniaxial pressure cells. Our setup exploits the ability of piezoelectric actuators to generate finite-frequency stresses and strains in the Hz-kHz range through the application of an a.c. voltage. Using the ruthenate Sr$_2$RuO$_4$ as a test-bed material for proof-of-principle measurements, we have shown that the Young's modulus data from our low-frequency a.c. setup are in very good agreement with data from static Young's modulus measurements. Our a.c. setup is well suited for detecting small anomalies in the strain dependence of the modulus. Furthermore, our a.c. data contains information on the phase between applied stress and resulting strain, which we find to be finite in the magnetic phase of Sr$_2$RuO$_4$ under high compression.

Our setup opens up the possibility to study the finite-frequency elastic response function in quantum materials which are subjected to time-varying external stress fields and large, static tuning stresses. This approach is akin to a.c. susceptibility studies \cite{Topping19acsusc,Grigera03}, where a time-dependent magnetic field acts as the driving external force to probe dynamics. Our method provides a new perspective on the viscoelastic behavior and lattice dynamics of solids. Following the fluctuation-dissipation theorem\cite{Kubo66}, the viscoelastic response is related to the low-frequency lattice dynamics of systems whenever they are amenable to stress tuning. It can be expected that the dynamics are governed by a range of intriguing phenomena, such as the movement of domain walls under strain, slow order-parameter dynamics and collective effects in general.



%
%

%

\begin{acknowledgments}
We thank You-Sheng Li and Clifford W. Hicks for highly useful insights from technical discussions. Financial support by the Max Planck Society is gratefully acknowledged. In addition, we gratefully acknowledge funding through the Deutsche Forschungsgemeinschaft (DFG, German Research Foundation) through TRR 288—422213477 (Project A10) and the SFB 1143 (project-id 247310070; project C09). Research in Dresden benefits from the environment provided by the DFG Cluster of Excellence ct.qmat (EXC 2147, project ID 390858490). C.I.O and Z.H. acknowledges the support of a St Leonards scholarship from the University of St. Andrews. N.K. work is supported by JSPS KAKENHI (No. JP18K04715, No. JP21H01033, and No. JP22K19093).
\end{acknowledgments}


\section{Appendix: Determination of the spring constant of the cell}
\label{sec:app1}


We follow the report of Barber \textit{et al.} \cite{Barber19Piezo} to determine the spring constant of our cell, which is similar in design, but not identical, to theirs. For the finite-element simulations, we used the software COMSOL \cite{COMSOL}. Since the cell is made out of titanium, we use the room-temperature Young's modulus of 103\,GPa and a Poisson's ratio of 0.33 for the simulations.

The spring constant of the cell contains various contributions (see Fig.\,\ref{fig:pressure-cell}): (i) the spring constant of the piezoelectric actuators, $k_\textrm{p,tot}$, (ii) the spring constant of the moving block B (the force block), $k_B$, and (iii) the spring constant of the moving block A, $k_{A}$.

According to the data sheet of the actuators \cite{PICMAactuators}, the spring constant of a single stack of type P-885.51 used here is $k_p\,=\,50\,$N$/\mu$m at room temperature. Combining the two compression actuators in parallel to each other and in series with the tension actuator gives $k_\textrm{p,tot}\,=\,33\,$N/$\mu$m.


To simulate $k_B$, we apply a force of 10\,N to moving block B at the position of the carrier mounting holes and measure its displacement. The resulting spring constant in the simulations was $k_{B}\,=\,(7.6\,\pm\,0.3)$\,N/$\mu$m. We also determined this spring constant experimentally, by hanging various weights from block B and measuring the resulting displacement. This procedure yielded an experimental value of $(7.7\,\pm\,0.1)$\,N/$\mu$m, which is, within the error bars, consistent with the simulated value.


The moving block A is designed to have a small spring constant in the direction of the applied force and a large spring constant in the orthogonal directions, to avoid torque on the sample. These rotational spring constants dominate $k_A$. For the simulations of $k_A$, we consider the application of force of 10\,N to an infinitely stiff sample. Since the sample is infinitely stiff, there is zero displacement between the carrier mounting points on block A and B ($\Delta d\,=\,0$) upon the application of 10\,N. Due to the finite rotational spring constant of block A, the actuators must apply a slightly higher force than 10\,N. Our simulations showed that a force of 10.45\,N is needed. We then evaluated the difference in actual displacement between the carrier mounting holes on block A and the actuator attachment area. Note that, even though there is zero relative displacement across the gap, block A is still displaced because block B moves under applied force (see Ref.\,\citen{Barber19Piezo} for further details on this simulation). The present simulation gives an estimate of $k_A\,=\,(7.7\pm2)$\,N/$\mu$m.


Taking $k_\textrm{p,tot}$, $k_A$ and $k_B$ together in series yields $k_\textrm{cell} = (3.4\,\pm\,0.5)$\,N/$\mu$m at room temperature.




%

\end{document}